\documentclass[amsmath,superscriptaddress,showpacs,prl,twocolumn]{revtex4-1}
\usepackage{bbm}
\usepackage{bm}
\usepackage{enumerate}
\usepackage{graphicx}
\usepackage[dvips]{epsfig}
\usepackage{epsf}
\usepackage[latin1]{inputenc}
\usepackage{amsmath}
\usepackage{amsfonts}
\usepackage{amssymb}
\usepackage[abs]{overpic}
\usepackage{xcolor}

\bibpunct{[}{]}{;}{n}{}{}

\begin{document}
		\title{Magnon Landau Levels and Spin Responses in Antiferromagnets}
	
\author{Bo Li}
\affiliation{Department of Physics and Astronomy and Nebraska Center for Materials and Nanoscience, University of Nebraska, Lincoln, Nebraska 68588, USA}

\author{Alexey A. Kovalev}
\affiliation{Department of Physics and Astronomy and Nebraska Center for Materials and Nanoscience, University of Nebraska, Lincoln, Nebraska 68588, USA}
	
\date{\today}

\begin{abstract}
We study gauge fields produced by gradients of the Dzyaloshinskii-Moriya interaction and propose a model of AFM topological insulator of magnons.
In the long wavelength limit, the Landau levels induced by the inhomogeneous Dzyaloshinskii-Moriya interaction exhibit relativistic physics described by the Klein-Gordon equation.
The spin Nernst response due to formation of magnonic Landau levels is compared to similar topological responses in skyrmion and vortex-antivortex crystal phases of AFM insulators. Our studies show that AFM insulators exhibit rich physics associated with topological magnon  excitations.

\end{abstract}
	\maketitle

Emergent electromagnetism~\cite{Volovik_1987,Tatara2008} is at the core of a multitude of fascinating physical phenomena ranging from topological Hall effect~\cite{PhysRevLett.93.096806,PhysRevLett.102.186602,PhysRevLett.102.186601,PhysRevLett.106.156603,Schulz2012,PhysRevB.95.064426,Gobel2017} in skyrmion crystals~\cite{Bogdanov.HubertJMMM1994,Roesler.Bogdanov.eaN2006,Muehlbauer.Binz.eaS2009,Yu.Onose.eaN2010} to formation of topological magnons~\cite{PhysRevB.87.174427,PhysRevB.87.144101,PhysRevB.90.024412,PhysRevLett.117.217203,Owerre2016,PhysRevB.96.224414,PhysRevB.97.174413,PhysRevB.97.134411,PhysRevB.98.060404}. Many applications related to information storage and processing can emerge from such useful features of magnetic systems as topological protection and low-dissipation spin transport~\cite{Fert2017,Zhou2014,Chumak2015,Gobel2019a}. The need for minimizing losses due to Joule heating has shifted the focus of recent research to insulating materials lacking itinerant electrons but still capable of carrying spin currents~\cite{RevModPhys.90.015005}.

Recently, antiferromagnets (AFM) became the focus of active research as they possess unique features associated with lack of stray fields and ultrafast dynamics in THz range~\cite{Olejnk2018}. Many spintronics concepts readily extend to AFM materials as is the case with spin-orbit torques~\cite{RevModPhys.91.035004} demonstrated experimentally in CuMnAs~\cite{PhysRevLett.113.157201,Wadley2016}. Skyrmions in AFM can be potentially stabilized by staggered fields~\cite{PhysRevB.96.060406,PhysRevB.100.100408} induced by field-like spin-orbit torques in CuMnAs and Mn$_2$Au or by coupling to boundary magnetization in Cr$_2$O$_3$. AFMs are expected to exhibit interesting physics associated with vanishing topological and skyrmion Hall effects~\cite{PhysRevB.92.020402,PhysRevLett.116.147203,Velkov_2016,Jin2016,Zhang2016}. 
The topological spin Hall effect in AFMs has been predicted for conducting systems~\cite{PhysRevB.96.060406,PhysRevLett.121.097204,pssr.201700007}. In insulating materials, the topological spin Hall effect mediated by magnons has been studied for isolated skyrmions~\cite{PhysRevB.99.224433}. The topological spin Nernst effect in skyrmion crystals have not been studied in insulators where the response can be associated with appearance of Landau levels of magnons~\cite{PhysRevB.87.024402,PhysRevLett.122.057204}.

In this Letter, we study gauge fields produced by gradients of the Dzyaloshinskii-Moriya interaction (DMI) and show that such fields can lead to realizations of magnon Landau levels and AFM magnonic topological insulator. In contrast to previous proposals ~\cite{PhysRevB.96.224414,Owerre2018}, in the long wavelength limit the proposed AFM magnonic topological insulator maps to the Klein-Gordon equation in the presence of uniform magnetic field and does not rely on the Aharonov-Casher effect with prefactor $1/c^2$, as gauge fields originate in DMI gradients.  The resulting Landau levels lead to unconventional steps in the accumulation of the spin Chern number and can be probed by measuring the spin Nernst response. We further compare such response to the magnonic topological spin Nernst effect in AFM skyrmion crystals and square crystals of vortices and antivortices. We confirm that the topological spin responses of AFM skyrmions can be qualitatively understood by considering Landau levels induced by a uniform magnetic flux; however, we also identify differences.   

\textit{AFM magnons and fictitious gauge fields}---We begin by implementing various gauge fields into description of AFM magnons. We consider magnonic excitations on top of a smooth N\'eel texture and in the presence of slowly varying DMI.
We consider the free energy density,  $\mathcal{F}[\boldsymbol{m},\boldsymbol{n}]=\mathcal{F}[\boldsymbol{n}]+\frac{\mathcal A}{2}\boldsymbol{m}^2$ with $\mathcal A$ being the inverse of the transverse spin susceptibility,  and replace $\boldsymbol{m}, \boldsymbol{n}$ by $\boldsymbol{m}=(\boldsymbol{m}_A+\boldsymbol{m}_B)/2$ and $\boldsymbol{n}=(\boldsymbol{m}_A-\boldsymbol{m}_B)/2$ where the sublattice spin fields are $\boldsymbol{m}_A$ and $\boldsymbol{m}_B$.
We also define:
\begin{equation}\label{FreeEnergy}
{\mathcal F}[\boldsymbol n]=\frac{\mathcal J}{2}\left(\partial_{i}\boldsymbol n\right)^{2}+\mathcal K (\boldsymbol n \cdot \hat{z})^{2}-\mathcal H_s (\boldsymbol n \cdot \hat{z})+ \boldsymbol{\mathcal{D}}_j (\partial_j \boldsymbol n \times \boldsymbol n),
\end{equation}
where we sum over repeated index $i=x,y$, $\boldsymbol n$ is a unit vector along the N\'eel order, $\mathcal{J}$ is the exchange constant, $\mathcal K$ is the effective uniaxial anisotropy, $\mathcal H_s$ is the staggered magnetic field arising due to the spin-orbit torque or the effect of boundary magnetization~\cite{PhysRevB.96.060406,PhysRevB.100.100408}, and $\mathcal D_{ij}=(\boldsymbol{\mathcal{D}}_j)_i$ is the DMI described by a general tensor.  We concentrate on the axially symmetric interface with a heavy metal for which there are only two non-zero tensor coefficients $\mathcal D_{12}=-\mathcal D_{21}=\mathcal D$ \cite{Kovalev2018}.

We assume that in the ground state $\boldsymbol{m}_0=0$ and $\boldsymbol{n}_0=(\sin\theta\cos\phi,\sin\theta\sin\phi,\cos\theta)$ where $\theta,\phi$ are spherical angles. This assumption ensures decoupling of the two chirality subspaces. Numerically, we see that lifting this assumption does not substantially modify our conclusions. The local spin field can be conveniently parametrized by a rotational matrix $R=\exp(L_z\phi)\exp(L_y\theta)$ with $(L_i)_{jk}=-\epsilon_{ijk}$ ($i=x,y,z$ or $1,2,3$) being the generators of rotational matrices. Specifically, $\boldsymbol{m}_{A(B)}=R \boldsymbol{m}_{A(B)}^\prime$, with $\boldsymbol{m}_A^\prime=\hat{z}\sqrt{1-|\gamma_A|^2}+\hat{x}\gamma_A^x+\hat{y}\gamma_A^y$ and 
 $\boldsymbol{m}_B^\prime=-\hat{z}\sqrt{1-|\gamma_B|^2}+\hat{x}\gamma_B^x-\hat{y}\gamma_B^y$, where $\gamma^{x,y}_{A(B)}$ stands for the spin wave, and $|\gamma_{A(B)}|^2=(\gamma_{A(B)}^x)^2+(\gamma_{A(B)}^y)^2$. We consider slowly varying spin textures and DMI and limit the discussion to the leading order of its spatial derivative. As the size of DMI induced textures scales as $\mathcal J/\mathcal D$, we systematically perform analysis up to the first order in $\mathcal D/\mathcal J$ and discard anisotropy and staggered magnetic field terms, assumed to be small when texture is present~\cite{PhysRevB.87.024402,PhysRevB.100.100408}. Plugging the rotational-matrix-parametrized spin field into the free energy $\mathcal{F}[\boldsymbol{m},\boldsymbol{n}]$ generates a Hamiltonian, in which magnons couple to a spin texture induced emergent gauge field ``$\boldsymbol{a}$" \cite{Kovalev2012,PhysRevB.98.134450,PhysRevLett.122.057204}, $\mathcal{H}_{mag}=\frac{1}{2}\psi^\dagger\hat{\mathcal{H}}\psi$ with $\hat{\mathcal{H}}=\hat{\mathcal{H}}_+\oplus \hat{\mathcal{H}}_-$,
 \begin{eqnarray}\label{eq:chiralH}
\hat{\mathcal{H}}_{\chi}= [\frac{\mathcal A}{8}-\frac{\mathcal J}{8}(\vec{\nabla}-i\chi\boldsymbol{a})^2]+[\frac{\mathcal A}{8}+\frac{\mathcal J}{8}(\vec{\nabla}-i\chi\boldsymbol{a})^2]\tau_1.
\end{eqnarray}Here, $\psi=(\psi_A,\psi^\ast_B,\psi_A^\ast,\psi_B)^T$ with $\psi_{A(B)}=\gamma_{A(B)}^x+i\gamma_{A(B)}^y$, $\tau_1$ is the Pauli matrix, $\chi=\pm 1$ labels the chirality of magnons. The emergent gauge field has two contributions,
$\boldsymbol{a}=\boldsymbol{a}^t+\boldsymbol{a}^d$, where $a^t_i=\cos\theta\partial_i\phi$ and $\boldsymbol{a}^d=-(\mathcal D/\mathcal J)\exp{(\pi L_z/2)}\boldsymbol{n}_0$. These two parts result in emergent magnetic fields, 
$b^t_i=(\vec{\nabla}\times \boldsymbol{a}^t)_i=-\frac{1}{2}\epsilon_{ijk}\boldsymbol{n}_0\cdot(\partial_j\boldsymbol{n}_0\times\partial_k\boldsymbol{n}_0)$, and
$\boldsymbol{b}^d=\vec{\nabla}\times \boldsymbol{a}^d$ (see details in the Supplemental
Material (SM)~\cite{Note}). The latter can generate an emergent magnetic field through an inhomogeneous DMI in the absence of spin textures. The in-plane component, $\boldsymbol a^d_{\parallel}=(\mathcal D/\mathcal{J})\boldsymbol n_0\times\hat z$, induces a fictitious magnetic field $\boldsymbol b^d=-\hat z(\vec\nabla \mathcal D\cdot\boldsymbol n_0)/\mathcal J$ (e.g. for $\mathcal D/ \mathcal J= B y$ and $\boldsymbol n_0=\hat{y}$ we get $\boldsymbol b^d=-B\hat{z}$). 

The kinetic term of magnons can be extracted from the Berry phase Lagrangian of spins \cite{auerbach1998interacting}, we obtain $\mathcal{L}_{kin}=i \mathcal{S} \psi^\dagger\sigma_3\otimes\tau_3\dot{\psi}/4$ with $\mathcal{S}$ being the spin density. The total Lagrangian density of magnon field is block-diagonal with respect to subspace $\eta_+=(\psi_A,\psi^\ast_B)^T$, $\eta_-=(\psi_A^\ast,\psi_B)^T$. The decoupled matrix Schr\"{o}dinger equations are 
\begin{eqnarray}\label{eq:Schrodinger}
i\chi \frac{\mathcal{S}}{2}\tau_3\partial_t\eta_\chi=\hat{\mathcal{H}}_\chi\eta_\chi.
\end{eqnarray}
We first consider the uniform emergent magnetic field which can be justified for uniform DMI gradient or smooth enough textures. 
In the Landau gauge, $\boldsymbol{a}_0=(yB,0,0)$, the eigenenergies are chirality degenerate, $\varepsilon_n^{\pm}= \pm\sqrt{\mathcal{J}\mathcal{A}B(2n+1)}/(2\mathcal{S})$, which agrees with Landau levels of the Klein-Gordon equation \cite{Lam1971}. The wave function can be found by substituting $\varphi_{n k_x}^\chi(\boldsymbol r)=(\alpha_1,\alpha_2)^T\xi_{n k_x}^\chi(\boldsymbol r)$ into Hamiltonian \eqref{eq:chiralH} where $\xi_{n k_x}^\chi(\boldsymbol r)$ is the known eigenfunction of $n$-th nonrelativistic Landau level~\cite{landau1977quantum}. The number of degenerate states is determined by the total number of the magnetic flux quanta.
The two species of magnons with opposite chirality feel opposite magnetic flux in Eq.~\eqref{eq:chiralH} as they are time-reversal partners of each other, which results in vanishing thermal Hall response (see SM~\cite{Note}). On the other hand, spin and chirality current responses are nonzero. 
\begin{figure}[!ht]
\centering
\includegraphics[width=\linewidth]{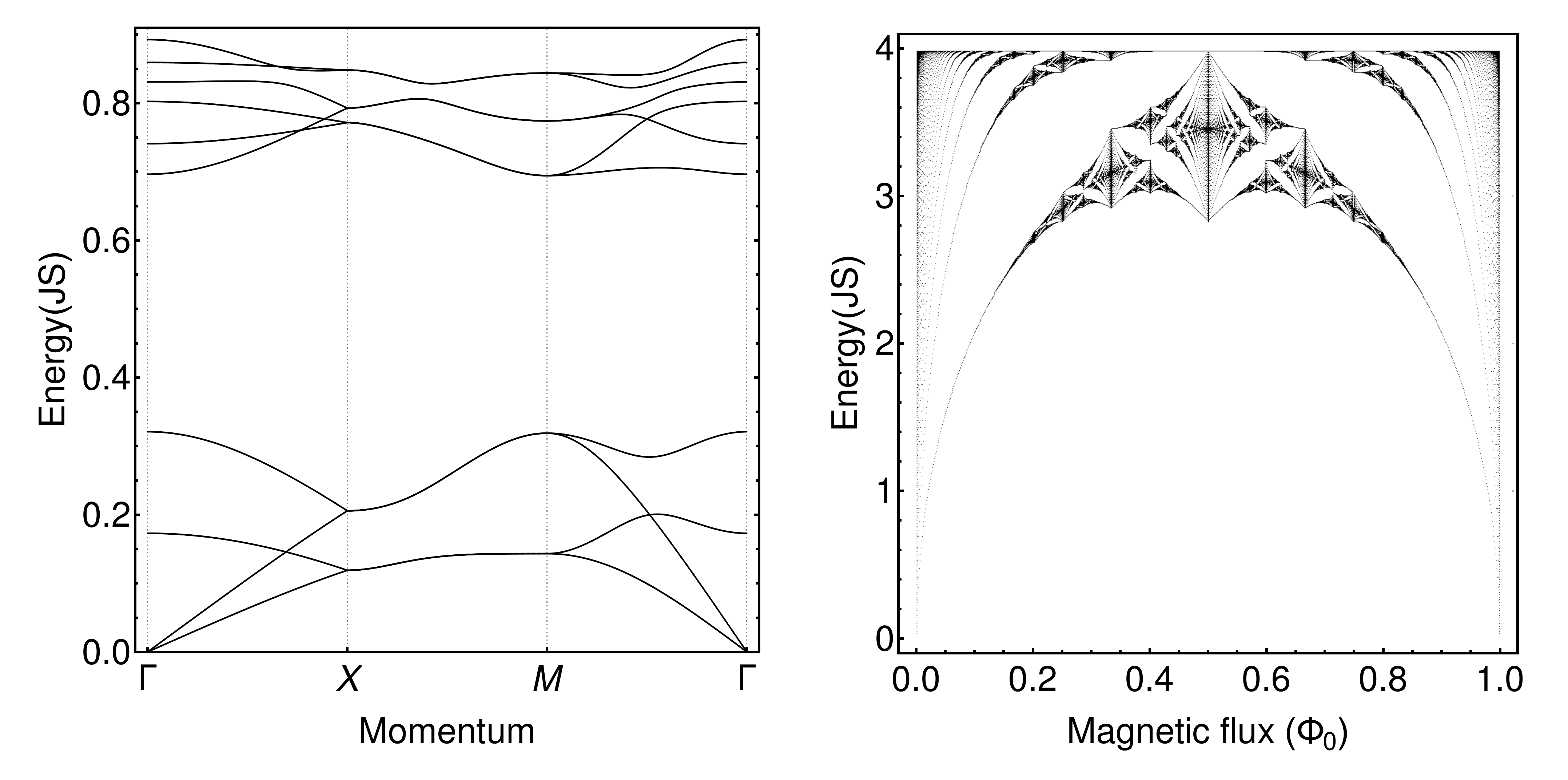}
\caption{Left: Lowest magnon bands of skyrmion crystal in a square lattice AFM along the Brillouin zone loop $\Gamma-X-M-\Gamma$. A splitting of chiral modes can be clearly identified. Right: The Hofstadter's butterfly of AFM with uniform magnetic flux $\Phi=\frac{p}{q}\Phi_0$ per unit cell for $q=1000$, $\Phi_0$ is the flux quantum.}
\label{fig:spectrum}
\end{figure}

\textit{Spin Nernst effect in AFM topological insulator}---In the absence of spin texture, Eq.~\eqref{eq:chiralH} can describe AFM topological insulator. The gauge field is induced by gradient of DMI and index $\chi$ also corresponds to the conserved spin $s_z$. To describe the magnonic topological insulators numerically, we construct and analyze lattice models of both FM and AFM with gradient of DMI (see SM~\cite{Note}). A square lattice Hamiltonian of collinear FM (AFM) reads
\begin{eqnarray}
H=\sum_{\langle ij\rangle}J\boldsymbol{S}_i\cdot\boldsymbol{S}_j+\boldsymbol{D}_{ij}(\boldsymbol{S}_i\times\boldsymbol{S}_j)-\sum_{i}H_i S^y_i-K ({S}_i^y)^2.
\end{eqnarray}
The order parameter is oriented along the y-axis to realize the Landau gauge. Above, the exchange parameter is $J<0$ ($J>0$) for FM (AFM), $H_i$ is (staggered) magnetic field, $K$ is the magnetic anisotropy, and $\boldsymbol{D}_{ij}=D(\boldsymbol r)\hat{z}\times \boldsymbol\delta_{ij}$ describes DMI with Rashba symmetry for a bond $\boldsymbol\delta_{ij}$. In FM case, we write the exchange and DMI terms in a rotated frame with the quantization axis along the y-axis as $\tilde J_{ij} (e^{-i\phi_{ij}} S_i^- S_j^++e^{i\phi_{ij}} S_i^+ S_j^-)/2+J S_i^z S_j^z$ where $\tilde J_{ij} e^{i\phi_{ij}}=J+i \boldsymbol{D}_{ij}\cdot \boldsymbol n_0$ with $\boldsymbol n_0$ being the direction of the order parameter~\cite{PhysRevLett.95.057205}. In AFM case, we need to replace $S_j^\pm \rightarrow S_j^\mp$, and $S_j^z \rightarrow -S_j^z$ for one of sublattices.

To replicate the Landau gauge, we assume that bonds are along the Cartesian coordinates and the strength of DMI is nonuniform, i.e., $D(\boldsymbol r)\delta/J=\tan[\delta B y]$ where $\delta$ is the bond length (when DMI is small $D(\boldsymbol r)/J\approx B y$, see details in SM~\cite{Note}). Using the Holstein-Primakoff transformation in the limit of large $S$, i.e. $S_j^+\approx \sqrt{2S} a_i$, $S_j^-\approx \sqrt{2S} a^\dagger_i$, $S_i^z\approx S-a^\dagger_i a_i$, we recover discreet realization of noninteracting magnons subjected to uniform magnetic field with a vector potential $\boldsymbol a_0=(yB,0,0)$. In the long wavelength limit, FM magnons are described by the Schr\"odinger equation while AFM magnons by the Klein-Gordon equation. We concentrate on AFM using FM system only for comparison, where in both cases the spin along the quantization axis is conserved.  
After the Fourier transform, the Hamiltonian for $s_z=1$ becomes  
\begin{equation}\label{eq:ham}
H_+=\frac{1}{2}JS\sum_{\boldsymbol k} \Psi_+^\dagger(\boldsymbol{k})\hat{H}_+(\boldsymbol k)\Psi_+(\boldsymbol{k}),
\end{equation}
where $\Psi_+=(a_1(\boldsymbol{k}),b^\dagger_1(-\boldsymbol{k})\dots b^\dagger_{2N}(-\boldsymbol{k}),a_{2N}(\boldsymbol{k}))^T$ is the bosonic field, and the unit cell contains $N$ by $2$ array of atoms from each sublattice of the square-lattice AFM. The Hamiltonian has a block structure
\begin{align}
\hat H_+(\boldsymbol k) = 
\begin{pmatrix}
\hat a & \hat b \\
\hat b & \hat a
\end{pmatrix} \,,
\end{align}
where for $2N\times2N$ matrices $\hat a$ and $\hat b$ the nonzero elements are given by $a_{i,j}=4$, $b_{i,j}=\cos(k_x+ j\phi_0)$ for $i=j$, and $a_{i,j}=a^*_{j,i}=e^{-i k_y}$ for $i-j=1$ modulo $2N$. Here the phase factor $\phi_0=2\pi p/q$ describes the strength of magnetic field, i.e., $2p$ is the number of flux quanta for enlarged unit cell and $q=2N$. For $s_z=-1$, $\hat H_-(\boldsymbol k)=\hat H^T_+(-\boldsymbol k)$ and $\Psi_-(\boldsymbol{k})=(a^\dagger_1(-\boldsymbol{k}),b_1(\boldsymbol{k})\cdots b_{2N}(\boldsymbol{k}),a^\dagger_{2N}(-\boldsymbol{k}))^T$.
The total Hamiltonian matrix can be diagonalized by a paraunitary matrix $T_{\boldsymbol{k}}$, i.e., $T_{\boldsymbol{k}}^\dagger \hat{H}T_{\boldsymbol{k}}=\hat{\mathcal{E}}_{\boldsymbol{k}}$, where $\hat{\mathcal{E}}_{\boldsymbol{k}}$ is a diagonal matrix describing eigenvalues~\cite{Colpa1978}. By varying strength of DMI, we can control the magnetic flux per unit cell which allows us to observe the Hofstadter's butterfly in full analogy with electronic systems (see Fig.~\ref{fig:spectrum}). Similarly to electronic systems, the exact energy bands can be found from expansion of $p/q$ into continuous fractions or from the Diophantine equation~\cite{PhysRevB.14.2239,PhysRevLett.49.405}. As can be seen from Fig.~\ref{fig:spectrum}, the form of the Hofstadter's butterfly differs from the case of nonrelativistic electrons.

In (non)collinear systems, the spin responses can be described by the spin Berry curvature~\cite{PhysRevLett.117.217203,PhysRevResearch.2.013079},
\begin{eqnarray}\label{eq:BerryCurvature}
\boldsymbol\Omega_{n}^\alpha=i\sum_{m\neq n}(\Tilde{\sigma}_3)_{nn}(\Tilde{\sigma}_3)_{mm}\frac{\frac{1}{2}\{\hat{\boldsymbol{v}},\hat{\Sigma}^\alpha\}_{nm}\times
\boldsymbol{\hat{v}}_{mn}}{(\bar{\varepsilon}_{n,\boldsymbol{k}}-\bar{\varepsilon}_{m,\boldsymbol{k}})^2},
\end{eqnarray}
where we define the anticommutator $\{\hat{\boldsymbol{v}},\hat{\Sigma}^\alpha\}=\hat{\boldsymbol{v}}\Tilde\sigma_3\hat{\Sigma}^\alpha+\hat{\Sigma}^\alpha\Tilde\sigma_3\hat{\boldsymbol{v}}$, $\bar{\varepsilon}_{m,\boldsymbol{k}}=(\Tilde\sigma_3\hat{\mathcal{E}}_{\boldsymbol{k}})_{mm}$, and the Pauli matrix in the particle-hole space, i.e., $(\Tilde\sigma_3)_{mm}=1$ for particle-like states and $(\Tilde\sigma_3)_{mm}=-1$ for hole-like states.  
The magnon spin density operator along the $\alpha$-axis is given by $\Sigma^\alpha(\boldsymbol{r})=\frac{1}{2}\Psi^\dagger(\boldsymbol{r})\hat{\Sigma}^\alpha\Psi(\boldsymbol{r})$ where $\hat{\Sigma}^\alpha=-\sigma_0\otimes\text{Diag}(m_1^\alpha, \cdots, m_{M}^\alpha)$ with the Pauli matrix $\sigma_0$ describing the particle-hole space and $\boldsymbol m_i$  being the direction of magnetic moment at position $i$ in a unit cell of $M$ atoms~\cite{PhysRevResearch.2.013079}. 
We consider the spin Nernst response~\cite{PhysRevB.93.161106}, $\alpha_{xy}^s=k_B/V \sum_{\boldsymbol{k},n=1}^N c_1(g(\varepsilon_{n,\boldsymbol{k}}))\Omega^{(z)}_n(\boldsymbol{k})$ where $g(\varepsilon)=(e^{\varepsilon/T}-1)^{-1}$ is the Bose-Einstein distribution and $c_1(x)=(1+x)\ln(1+x)-x\ln(x)$.
Due to degeneracy, we apply Eq.~\eqref{eq:BerryCurvature} to each subspace $s_z=\pm1$ separately. The total spin Chern number is a sum of spin Chern numbers for each subspace,  i.e., $C_n^s=(1/2 \pi)\int_{BZ} \Omega^{(z)}_n d^2k$ where $\Omega^{(z)}_n=\Omega^{(z)+}_n+\Omega^{(z)-}_n$. 
\begin{figure}[!ht]
\centering
\includegraphics[width=0.9\linewidth]{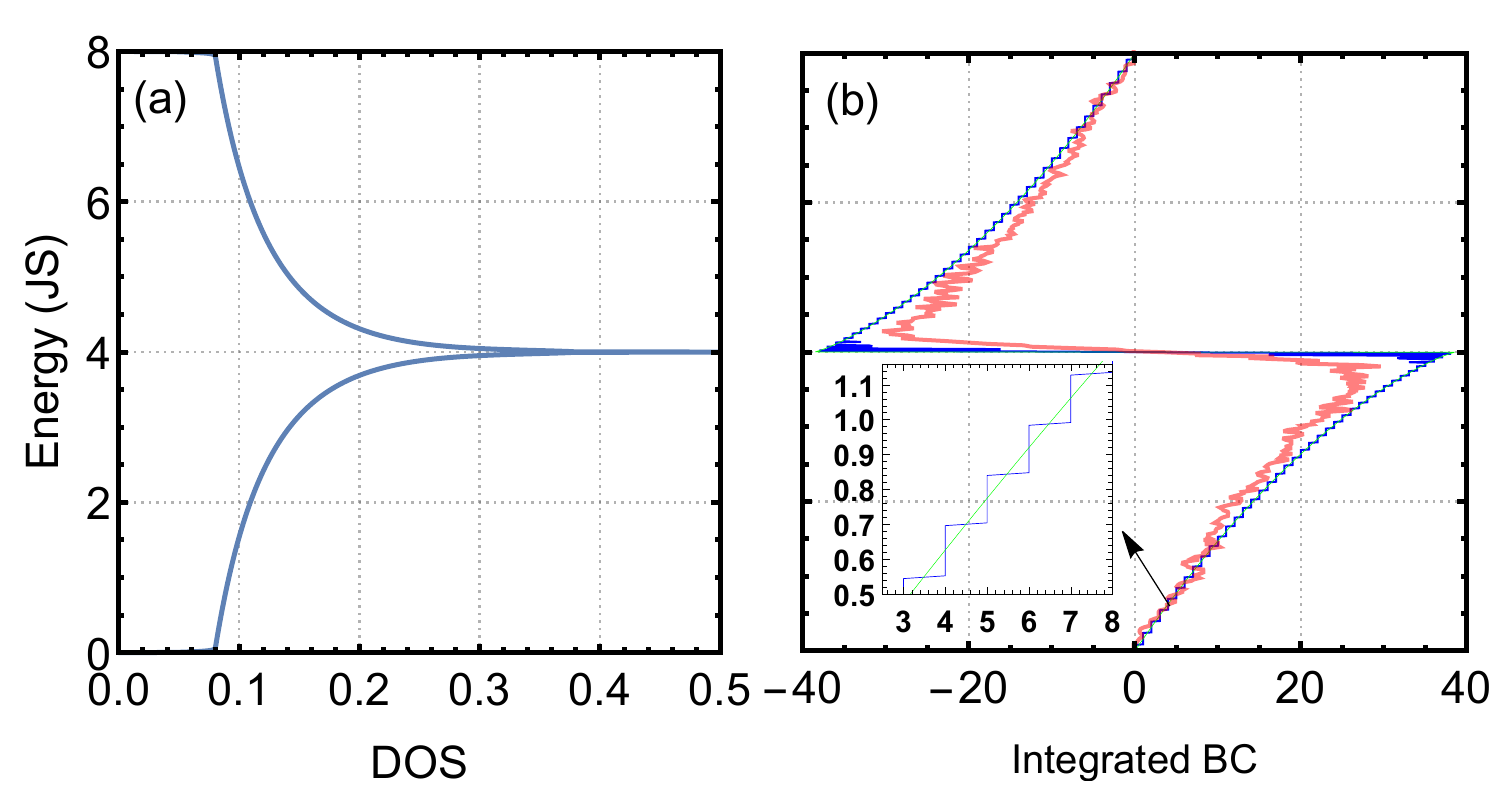}
\includegraphics[width=0.9\linewidth]{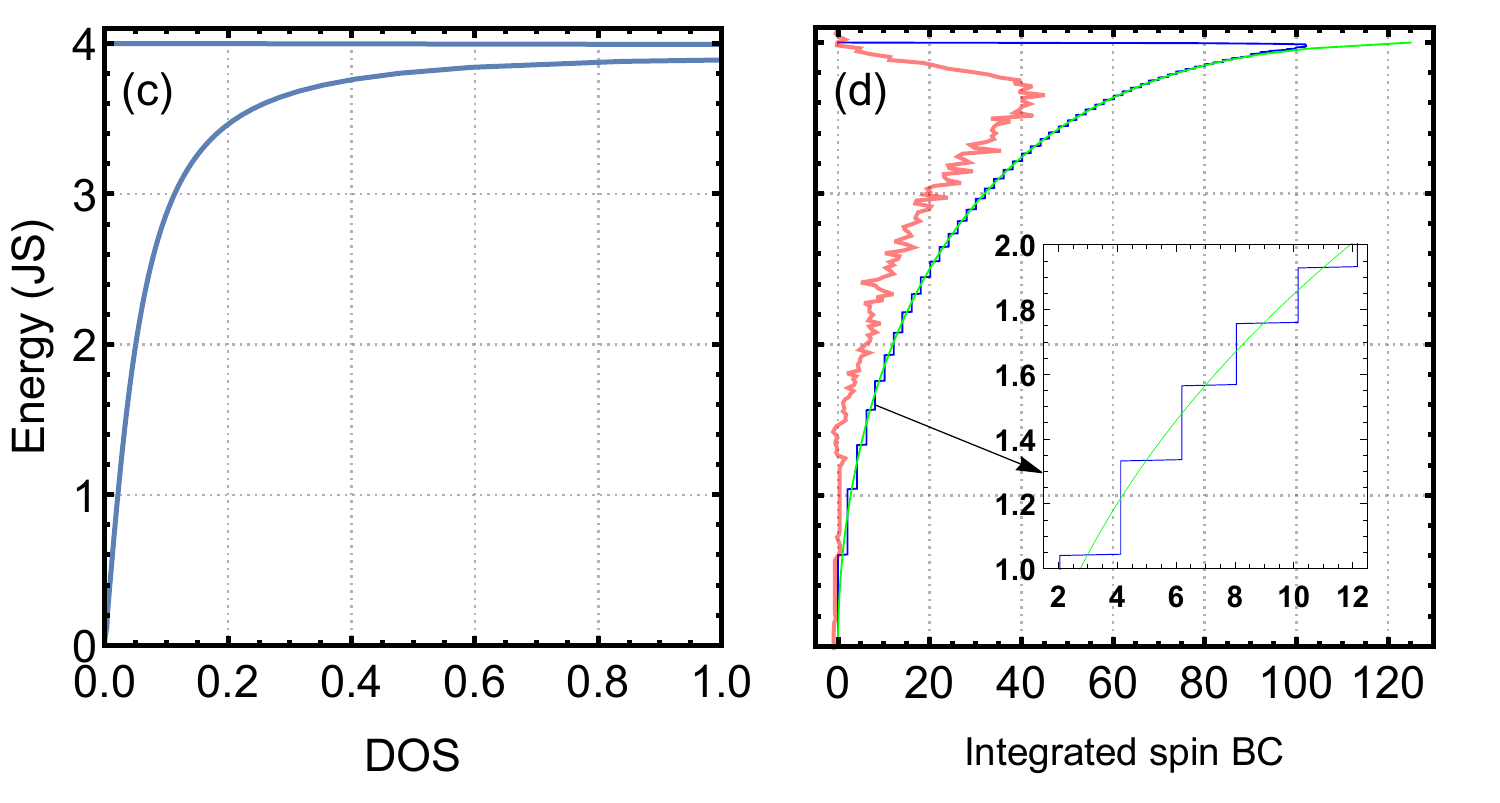}
\caption{(Color online) (a) and (c) The density of states (DOS) of magnons in a square lattice FM or AFM in the absence of gauge fields. (b)  The total (integrated) Berry curvature due to flux induced by DMI (blue curve) for $p=1$ and $q=77$. The same but nonuniform flux is produced by two skyrmions in SkX unit cell of $14\times22$ atoms for which the total Berry curvature is shown by red curve. (d) The total (integrated) spin Berry curvature due to flux induced by DMI (blue curve) for $p=2$ and $q=270$. The same but nonuniform flux is produced by two skyrmions in AFM SkX unit cell of $18\times30$ atoms for which the total sublattice Berry curvature is shown by red curve. In both plots the semiclassical approximation is shown by green curve.}
\label{fig:FM}
\end{figure}

\begin{figure}[!ht]
\centering
\includegraphics[width=0.6\linewidth]{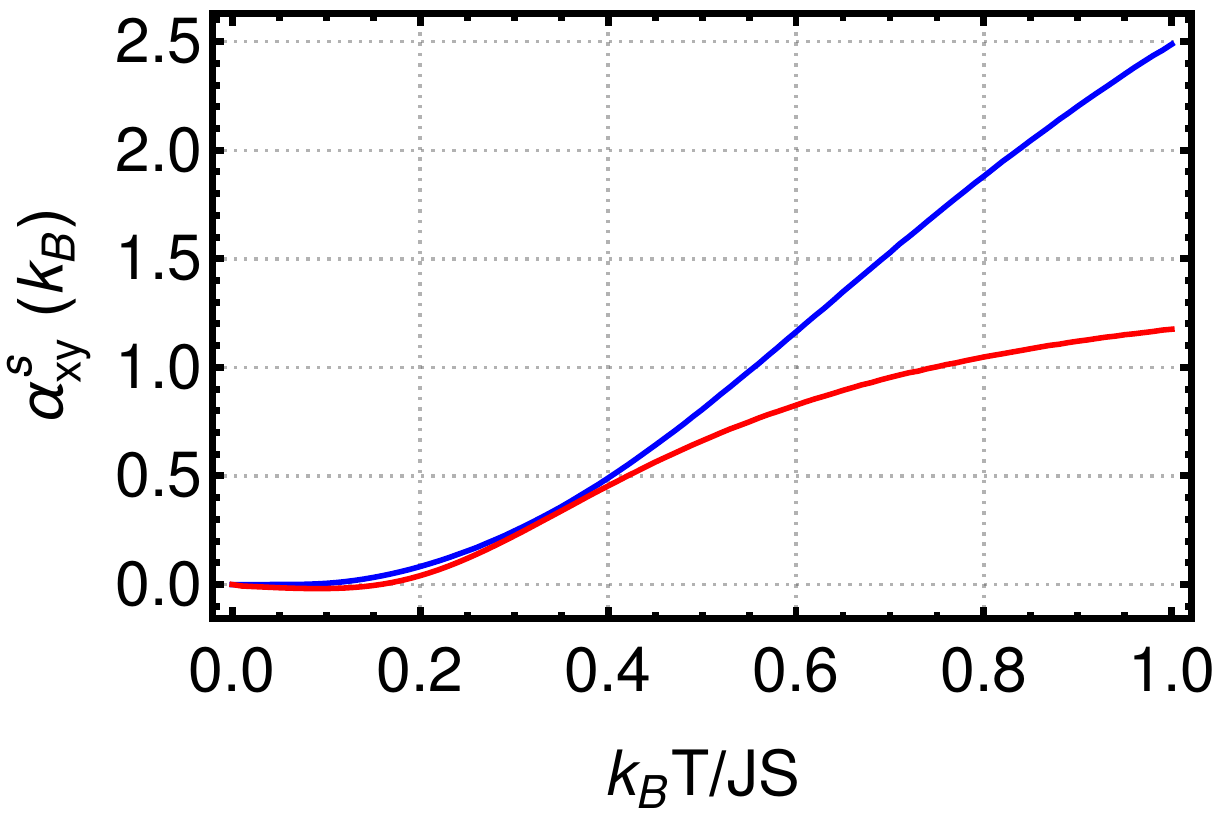}
\caption{Spin Nernst conductivity as a function of temperature. Red curve describes the topological spin Nernst response for square lattice AFM with a unit cell of $18\times30$ atoms containing two skyrmions. Blue curve describes the spin Nernst response in AFM magnonic topological insulator with DMI induced fictitious flux $\Phi=\frac{p}{q}\Phi_0$ for $p=2$ and $q=270$.}
\label{fig:SN}
\end{figure}

To establish a connection to the quantum Hall effect, we study the total Berry curvature of states below a certain energy, $C^s(\varepsilon)=(1/2 \pi) \int_{BZ} \sum_{\varepsilon_{n,k}<\varepsilon} \Omega^{(z)}_{n} d^2k$. For FM magnons, the results for the total Berry curvature and the magnon density of states (DOS) are shown in Figs.~\ref{fig:FM}(a) and (b). We observe a behavior associated with the van Hove singularity~\cite{PhysRevB.95.094413} of the magnon band structure. This causes a sign change in the total Berry curvature at the transition between particle- and hole-like states~\cite{PhysRevB.74.205414,PhysRevB.73.233406}. For AFM magnons, the total spin Berry curvature shown in Fig.~\ref{fig:FM}(d) exhibits steps of $2$ and uneven energy height even in the long wavelength limit. We observe sharp change in the spin Berry curvature at the DOS singularity in Fig.~\ref{fig:FM}(c). For both FM and AFM magnons, away from DOS singularity the formation of magnon Landau levels can be described by the Onsager's quantization scheme~\cite{Kaganov,PhysRevB.79.075429}. We confirm this by comparing the semiclassical curve corresponding to the area enclosed by DOS with the Berry curvature curves in Fig.~\ref{fig:FM}. Finally, the spin Nernst response is shown in Fig.~\ref{fig:SN}.  

\begin{figure}[!ht]
\centering
\begin{overpic}[width=0.7\linewidth]{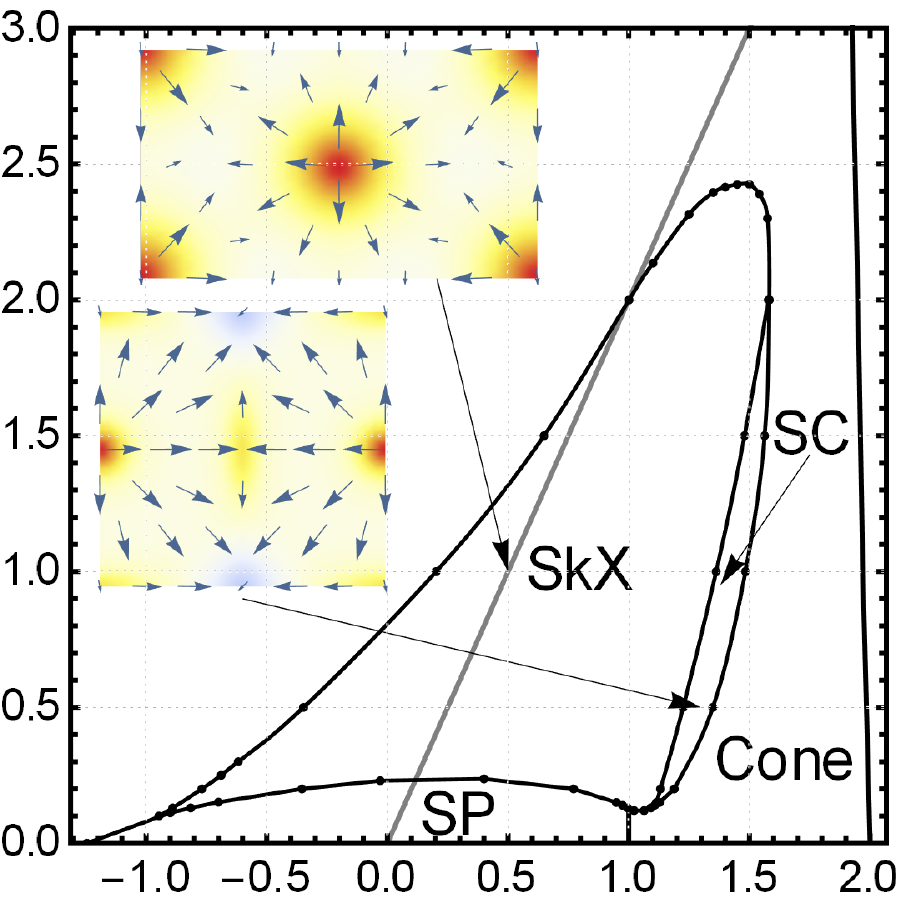}
\put(80,-10){$\mathcal{K J/D}^2$}
\put(-15,80){\rotatebox{90}{$\mathcal{J H}_s \mathcal{/D}^2$}}
\end{overpic}
\vspace{0.5cm}
\caption{(Color online) Zero temperature phase diagram of AFM with DMI. The axes correspond to the dimensionless staggered magnetic field and dimensionless effective anisotropy. The gray line separates the aligned and the tilted regions of the FM phase. This phase is taken over by the hexagonal skyrmion lattice (SkX), spiral (SP), cone phase, and the square crystal of vortices and antivortices (SC). The upper inset shows a hexagonal lattice unit cell with a skyrmion in the center. The lower inset shows a square crystal unit cell with AFM antimeron in the center. Red and yellow correspond to positive topological charge density and blue corresponds to the negative topological charge density.}
\label{fig:phase}
\end{figure}

\textit{AFM skyrmions and topological spin Nernst effect}--- 
The zero temperature phase diagram in Fig.~\ref{fig:phase} has been calculated by energy minimization~\cite{Vansteenkiste2014} from Eq.~\eqref{FreeEnergy} combined with rescaling of unit cell~\cite{Utkan2016}.
The free energy density in Eq.~\eqref{FreeEnergy} and the resulting phase diagram can also describe other spin textures obtained from N\'eel skyrmions by a global transformation in spin space (e.g. antiskyrmions or Bloch skyrmions) \cite{Utkan2016}. 
In addition to AFM-SkX phase identified in Ref.~\cite{PhysRevB.100.100408}, we also identify AFM-SC vortex-antivortex lattice~\cite{PhysRevB.80.054416,PhysRevB.91.224407,Utkan2016,Ozawa2016,Vousden2016,Yu2018} stabilized by the inplane anisotropy. Such textures can also contain antiferromagnetic antimerons with fractional topological charge as shown in Fig.~\ref{fig:phase}. In the absence of DMI gradients, we study the effect of fictitious magnetic fields where each SkX or SC unit cell with topological charge one contributes two flux quanta. 

For a uniform fictitious field approximation, $\boldsymbol{b}=-B\hat{z}$, with $B=|\langle\vec{\nabla}\times \boldsymbol{a}\rangle|=4\pi\langle\rho_{top}\rangle>0$, where $\rho_{top}=\boldsymbol{n}_0\cdot(\partial_x\boldsymbol{n}_0\times\partial_y\boldsymbol{n}_0)$.
This reproduces results from the previous section.
For a nonuniform fictitious field of skyrmion lattice with basis vectors $\vec{a}_1$ and $\vec{a}_2$, the Landau-level wave functions can be linearly combined to a new periodic basis for each energy level, $\tilde\varphi^\chi_{n m \boldsymbol k}$, which satisfies $T_{\vec{a}_{1(2)}}\tilde\varphi^\chi_{n m \boldsymbol k}=e^{i\boldsymbol{k}\cdot\vec{a}_{1(2)}}\tilde\varphi^\chi_{n m \boldsymbol k}$ with magnetic translational operator $T_{\vec{a}_{1,2}}$ satisfying $T_{\vec{a}_1}T_{\vec{a}_2}=e^{i\chi Q 4\pi}T_{\vec{a}_2}T_{\vec{a}_1}$. The phase factor indicates that each skyrmion unit cell contains topological charge $Q$ which leads to splitting into $2|Q|$ subbands described by quantum number $m$. In this new basis, one can include perturbations to Hamiltonian due to nonuniform fictitious flux and higher order terms disregarded earlier~\cite{PhysRevB.87.024402} (see SM~\cite{Note} for more details). This treatment leads to splitting of Landau levels and to coupling of magnons with opposite chiralities, as confirmed by calculating the magnon spectrum of skyrmion crystal in a square lattice AFM in Fig.~\ref{fig:spectrum}. 

To understand the effect of splitting of Landau levels, we study a square lattice AFM SkX and magnon excitations numerically.
Magnon excitations on top of textures in Fig.~\ref{fig:phase} can be described by the Holstein-Primakoff transformation in a local frame~\cite{PhysRevLett.122.187203}. The resulting Hamiltonian describes noninteracting magnons and can be diagonalized using the paraunitary matrices. Spectrum for the lowest bands of a lattice contaning $18\times30$ atoms is shown in Fig.~\ref{fig:spectrum}. We observe that the Landau levels become dispersive and that AFM chiral modes split. The total  sublattice Berry curvature is shown in Fig.~\ref{fig:FM}(d) where we use sublattice index instead of spin index in Eq.~\eqref{eq:BerryCurvature}. The sublattice index in Eq.~\eqref{eq:chiralH} and spin index in Eq.~\eqref{eq:ham} can be mapped onto each other in the absence of coupling between chiral modes. We observe only qualitative agreement with Landau levels in AFM calculated earlier for uniform flux due to coupling of chiral modes in AFM SkX and nonuniformity of flux. In Fig.~\ref{fig:FM}(b), we observe better agreement between Berry curvatures calculated for FM SkX in lattice of $14\times22$ atoms and for Landau levels in FM with uniform flux. The sign change of the Berry curvatures in Figs.~\ref{fig:FM}(b) and (d) can lead to the sign change of the topological thermal Hall and spin Nernst responses as a function of temperature.
Using the spin Berry curvature for the z-component of spin~\cite{PhysRevB.96.060406} (see SM~\cite{Note}), we calculate the topological spin Nernst response in Fig.~\ref{fig:SN} and confirm the sign change. As expected, the spin Nernst response in AFM SkX is suppressed compared to similar response in AFM topological insulator (see Fig.~\ref{fig:SN}). Note that at higher temperatures, a description relying on noninteracting magnons can become unreliable.

\textit{Conclusions}---We have constructed a model of AFM topological insulator of magnons. The fictitious flux is induced by inhomogeneous DMI and leads to formation of unconventional Hofstadter's butterfly. AFM magnon Landau levels exhibit large spin Nernst response and in the long wavelength limit are described by the Klein-Gordon equation. Landau levels characterized by energy scale $\sqrt{\mathcal A\mathcal J B}/\mathcal S \approx 0.4$~meV can be achieved by DMI change of $0.5$ mJ/m$^2$, e.g. in NiO/Au, over the length of $500$ nm~\cite{PhysRevLett.118.147201,Cao2018,Zhu2019,Lau2019,2020arXiv200711714A}. Similar physics also arises in AFM-SkX and AFM square vortex-antivortex phase leading to a topological spin Nernst response. This response is associated with the formation of dispersive Landau levels. Our predictions can be tested in magnetoelectrics with boundary magnetization~\cite{He2010}, rare earth garnet ferrimagnets, and AFM with DMI due to  structural asymmetry induced by neighbouring layer~\cite{Shao2019,2020arXiv200711714A}. 

This work was supported by the U.S. Department of Energy, Office of Science, Basic Energy Sciences, under Award No. DE-SC0014189.

\bibliographystyle{apsrev}
\bibliography{AFLandau}

\end{document}